\documentclass[doublecol]{epl2} 
\pdfoutput=1 %This is for the ArXiv submission only

\begin{document}

\title{Galaxy distribution and extreme value statistics}
\shorttitle{Galaxy distribution and extreme value statistics}

\author{Tibor Antal \inst{1} 
\and Francesco Sylos Labini\inst{2,3} 
\and Nikolay L. Vasilyev \inst{4} 
\and Yurij V. Baryshev \inst{4}
}
\shortauthor{Antal  \etal}

\institute{ \inst{1} Program for Evolutionary Dynamics, Harvard
  University, Cambridge, MA 02138 \\ \inst{2} Museo Storico della
  Fisica e Centro Studi e Ricerche Enrico Fermi, - Piazzale del
  Viminale 1, 00184 Rome, Italy \\ \inst{3} Istituto dei Sistemi
  Complessi CNR, - Via dei Taurini 19, 00185 Rome, Italy \\ \inst{4}
  Institute of Astronomy, St.Petersburg State University - Staryj
  Peterhoff, 198504, St.Petersburg, Russia }

\abstract{We consider the conditional galaxy density around each galaxy, and study its fluctuations in the newest samples of the Sloan Digital Sky Survey Data Release 7.  Over a large range of scales, both the average conditional density and its variance show a nontrivial scaling behavior, which resembles to criticality. The density depends,  for $10 \le r\le 80$ Mpc/h, only weakly (logarithmically) on the system size. Correspondingly, we find that the density fluctuations follow the Gumbel distribution of extreme value statistics. This distribution is clearly distinguishable from a Gaussian distribution, which would arise for a homogeneous spatial galaxy configuration. We also point out similarities between the galaxy distribution and critical systems of statistical physics.}

\maketitle

\section{Introduction}

One of the cornerstones of modern cosmology is the mapping of three
dimensional galaxy distributions. In the last decade two extensive
projects, the Sloan Digital Sky Survey (SDSS --- \cite{york}) and the
Two degree Field Galaxy Redshift Survey (2dFGRS --- \cite{colless03}),
have provided redshifts of an unprecedented quality for more than one
million galaxies.  A common feature observed in these surveys
\cite{busswell03,gott} is that galaxies are organized in a complex
pattern, characterized by large scale structures: clusters,
super-clusters, and filaments with large voids of extremely low local
density \cite{vogeley}. Recent analyses of these catalogs have shown
that galaxy structures display large amplitude density fluctuations at
all scales limited only by sample sizes
\cite{sdss_epl,sdss_aea,sdss_bao,2df_epl,2df_aea}.  In addition, the
conditional density \cite{book} has been found to decay with distance
as a power-law function with an exponent close to one, up to $\sim 30$
Mpc/h \footnote{We use $H_0=100h$ km/sec/Mpc, with $0.4\le h \le 0.7$,
  for the Hubble's constant.}. At larger scales, the situation was
unclear since in the 2dFGRS the relatively small solid angle prevents
the proper characterization of correlations at larger scales
\cite{2df_epl,2df_aea}. Conversely, the SDSS samples (data release 6
--- DR6) clearly show that conditional fluctuations are not
self-averaging for $r>30$ Mpc/h. In the latter case, the sample
volumes were found to be too small to obtain statistically stable
result due to wild fluctuations \cite{sdss_epl,sdss_aea}. Therefore,
although there are unambiguous evidences for the inhomogeneity of the
galaxy distribution at least up to scales of 100 Mpc/h
\cite{sdss_epl,sdss_aea,2df_epl,2df_aea}, the scaling properties at
scales larger than 30 Mpc/h were poorly understood.

The new galaxy samples from the data release 7 (DR7 --- \cite{dr7}) 
%of the SDSS 
doubled in size since the DR6 sample.
This new catalog is large enough to facilitate the study of
fluctuations in the galaxy distribution. In particular, we calculate
the galaxy density in a sphere of radius $r$ around each galaxy, {\it
  i.e.,} the conditional density. For uniformly positioned galaxies
\cite{book}, the average conditional density is independent of the
radius $r$, and the fluctuations over galaxies are
Gaussian. Conversely, in DR7 we find that the average density
depends logarithmically on $r$, while the fluctuations follow the
Gumbel distribution of extreme value statistics. This behavior has
an analog in statistical physics, where logarithmically changing
averages tend to correspond to Gumbel type fluctuations
\cite{bramwell09}.

  The rest of the paper is organized as follows. We first discuss the
  quantities we consider in the measurements and briefly discuss the
  main properties of the Gumbel distribution.  We then introduce the
  galaxy samples and our main results on the average, the variance and
  the fluctuation distribution of the conditional density we measured
  in the data. Finally we discuss the results and draw conclusions.

%%%%%%%%%%%%%%%%%%%%%%
\section{Statistical methods}

In this section we describe the estimators we use in the analysis and
then discuss the properties of the Gumbel distribution. We also
provide some physical examples where the Gumbel distribution was found
fit to experimental data.

\subsection{Estimators and their main properties}

A particularly useful characterization of statistical properties of
point distributions can be obtained by measuring conditional
quantities \cite{book}.  In this paper we focus on such a quantity,
namely we calculate the number $N(r)$ of galaxies contained in a
sphere of radius $r$ centered on a galaxy.  Note that not all galaxies
can be considered as sphere centers for a given radius $r$: a central
galaxy has to be farther than distance $r$ from any border of the
sample, so that the sphere volume is fully contained inside the sample
volume \cite{sdss_aea,2df_aea}. As $r$ approaches the radius of the
largest sphere fully contained in the sample volume, the statistics
become poorer.  To deal with these limitations for large values of
$r$, two effects should be taken into account: (i) the number of
points $M(r)$ satisfying the above condition is largely reduced and
(ii) most of the points are located in the same region of the sample.
Any conclusion about statistical properties must consider a careful
analysis of these limitations \cite{sdss_aea}.

%%%%%

\subsection{The Gumbel distribution} 

The Gumbel (also known as Fisher-Tippet-Gumbel) distribution is one of
the three extreme value distribution \cite{fisher28,gumbel58}. It
describes the distribution of the largest values of a random variable
from a density function with faster than algebraic (say exponential)
decay. The Gumbel distribution's PDF is given by
\begin{equation}
\label{gumbel}
 P(y)= \frac{1}{\beta} 
\exp\left[ - \frac{y-\alpha}{\beta} - 
\exp\left( - \frac{y-\alpha}{\beta} \right) \right] \;.
\end{equation}
With the scaling variable
\begin{equation}
\label{scalevar}
 x = \frac{y-\alpha}{\beta}
\end{equation}
the density function (Eq.\ref{gumbel}) 
simplifies to the parameter-free Gumbel
\begin{equation}
\label{gumbelscaled}
 P(x) = e^{-x-e^{-x}}
\end{equation}
with (cumulative) distribution $e^{-e^{-x}}$.  Note that this
distribution corresponds to large extremes, while for low extreme
values, $x$ is used instead of $-x$ in the Gumbel distribution.

The mean and the standard deviation (variance) of the Gumbel
distribution (Eq.\ref{gumbel}) is
\begin{equation}
\label{cumu}
 \mu = \alpha + \gamma \beta, \quad 
 \sigma^2 = (\beta\pi)^2/ 6 
\end{equation}
where $\gamma=0.5772\dots$ is the Euler constant. For the scaled
Gumbel (Eq.\ref{gumbelscaled}) the first two cumulants of Eq.\ref{cumu}
simplify to $\gamma$ and $\pi^2/6$.

%%%%%%%%%%%%%%%%%%%%%%%%%
\subsection{Gumbel in critical systems}

Away from criticality, any global (spatially averaged) observable of a
macroscopic system has Gaussian fluctuations, in agreement with the
central limit theorem (CLT). At criticality, however, the correlation
length tends to infinity, and the CLT no longer applies. Indeed,
fluctuations of global quantities in critical systems usually have
non-Gaussian fluctuations. The type of fluctuations is characteristic
to the universality class of the system's critical behavior
\cite{foltin94,racz02}.

To fit experimental data, the generalized Gumbel PDF
$P(x)=(e^{-x-e^{-x}})^a$ has often been used, where $a$ is a real
parameter. For integer values of $a$, this distribution corresponds to
the $a$-th maximal value of a random variable. The $a=1$ case
corresponds to the Gumbel distribution. Experimental examples for
Gumbel or generalized Gumbel distributions include power consumption
of a turbulent flow \cite{bramwell98}, roughness of voltage
fluctuations in a resistor (original Gumbel $a=1$ case)
\cite{antal01}, plasma density fluctuations in a tokamak
\cite{milligen05}, orientation fluctuations in a liquid crystal
\cite{joubaud08}, and other systems cited in \cite{bramwell09}.  The
Gumbel distribution describing fluctuations of a global observable was
first obtained analytically in \cite{antal01} for the roughness
fluctuations of $1/f$ noise. Its relations to extreme value statistics
have been clarified \cite{bertin05,bertin06}, generalizations have
appeared \cite{antal02}, and related finite size corrections have been
understood \cite{gyorgyi08}.

In a recent paper Bramwell \cite{bramwell09} conjectured that only
three types of distributions appear to describe fluctuations of global
observables at criticality. In particular, when the global observable
depends logarithmically on the system size, the corresponding
distribution should be a (generalized) Gumbel. For example the mean
roughness of $1/f$ signals depends on the logarithm of the observation
time (system size), and the corresponding PDF is indeed the Gumbel
distribution \cite{antal01}.

%%%%%%%%%%%%%%%%%%%%%%%%%%%%%%%%%%%%

\section{The Data}

We have constructed several sub-samples of the main-galaxy (MG) sample
of the spectroscopic catalog SDSS-DR7 \footnote{{\tt
    http://www.sdss.org/dr7}}.  We have constrained the flags
indicating the type of object to select only the galaxies from the MG
sample.  We then consider galaxies in the redshift range $10^{-4} \leq
z \leq 0.3$ with redshift confidence $z_{conf} \ge 0.35$ and with
flags indicating no significant redshift determination errors.  In
addition we apply the apparent magnitude filtering condition $m_r <
17.77$  \cite{strauss2002}.  The angular region we consider is
limited, in the SDSS internal angular coordinates, by $-33.5^{\circ}
\le \eta \le 36.0^\circ$ and $-48.0^\circ \le \lambda \le 51.5^\circ$:
the resulting solid angle is $\Omega=1.85$ steradians.  We do not use
corrections for the redshift completeness mask or for fiber collision
effects. Fiber collisions in general do not present a problem for
measurements of large scale galaxy correlations \cite{strauss2002}.
Completeness varies most near the current survey edges, which are
excluded in our samples.  In addition the completeness mask could be
the main source of systematic effects on small scale only, while we
are interested on the correlation properties on relatively large
separations \cite{sdss_bao}.

To construct volume-limited (VL) samples we computed the metric
distances $R$ using the standard cosmological parameters, {\it i.e.,}
$\Omega_M=0.3$ and $\Omega_\Lambda=0.7$. We computed absolute
magnitudes $M_r$ using Petrosian apparent magnitudes in the $m_r$
filter corrected for Galactic absorption.  We considered the sample
limited by $R\in[70,450]$ Mpc/h and $M_r\in [-21.8, -20.8]$ containing
$M=93821$ galaxies. In this sample there are about 1/5 of the whole
galaxies in DR7 it has relatively large spatial extensions and small
spread in galaxy luminosity.  Note that in other samples limited at
scales smaller than $\sim 400$ Mpc/h we found similar results.

We have checked that our main results in this VL sample do not depend
significantly on K-corrections and/or evolutionary corrections as
those used by \cite{blanton2003}. In this paper we use standard
K-correction from the VAGC data \footnote{{\tt
    http://sdss.physics.nyu.edu/vagc/}} (see discussion in
\cite{sdss_aea} for more details).

%%%%%%%%%%%%%%%%%%%%%%%%%%%%%%%%%%%%
\section{Results} 

In this section we present our findings from the analysis of the
galaxy data.

We have computed the number of galaxies $N_i(r)$ within radius $r$
around each galaxy $i$ satisfying the boundary condition previously
mentioned. By normalizing it by the volume $V_r=4\pi r^3/3$ of the
sphere, we obtain the conditional galaxy density $n_i(r)=N_i(r)/V$
around each galaxy $i$. This quantity is our main interest in this
paper.  The variable $n_i(r)$ differs for each galaxy, hence we
consider this local density $n_i(r)$ as a random variable, and study
its statistical properties. For example the conditional average
density within radius $r$ is defined as
\begin{equation}
  \label{eq:condden}
  \overline{n(r)} = \frac{1}{M(r)} \sum_{i=1}^{M(r)} n_i(r) \;,
\end{equation}
where $\overline{n(r)}$ is
``conditioned'' on the presence of the central galaxy.  
The simplest quantity to characterize density fluctuations is the
variance, or mean square deviation at scale $r$, which is defined as 
\begin{equation}
  \label{eq:condvar}
  \sigma^2(r) \equiv \mathrm{var}\, [n(r)] = \frac{1}{M(r)}
  \sum_{i=1}^{M(r)} n_i^2(r) - {\overline{n(r)}}^2 \;.
\end{equation}
In the following subsections we are going to study the whole
distribution of $n_i(r)$ as well.

\subsection{Self-averaging properties}

\begin{figure}
\onefigure[scale=0.32]{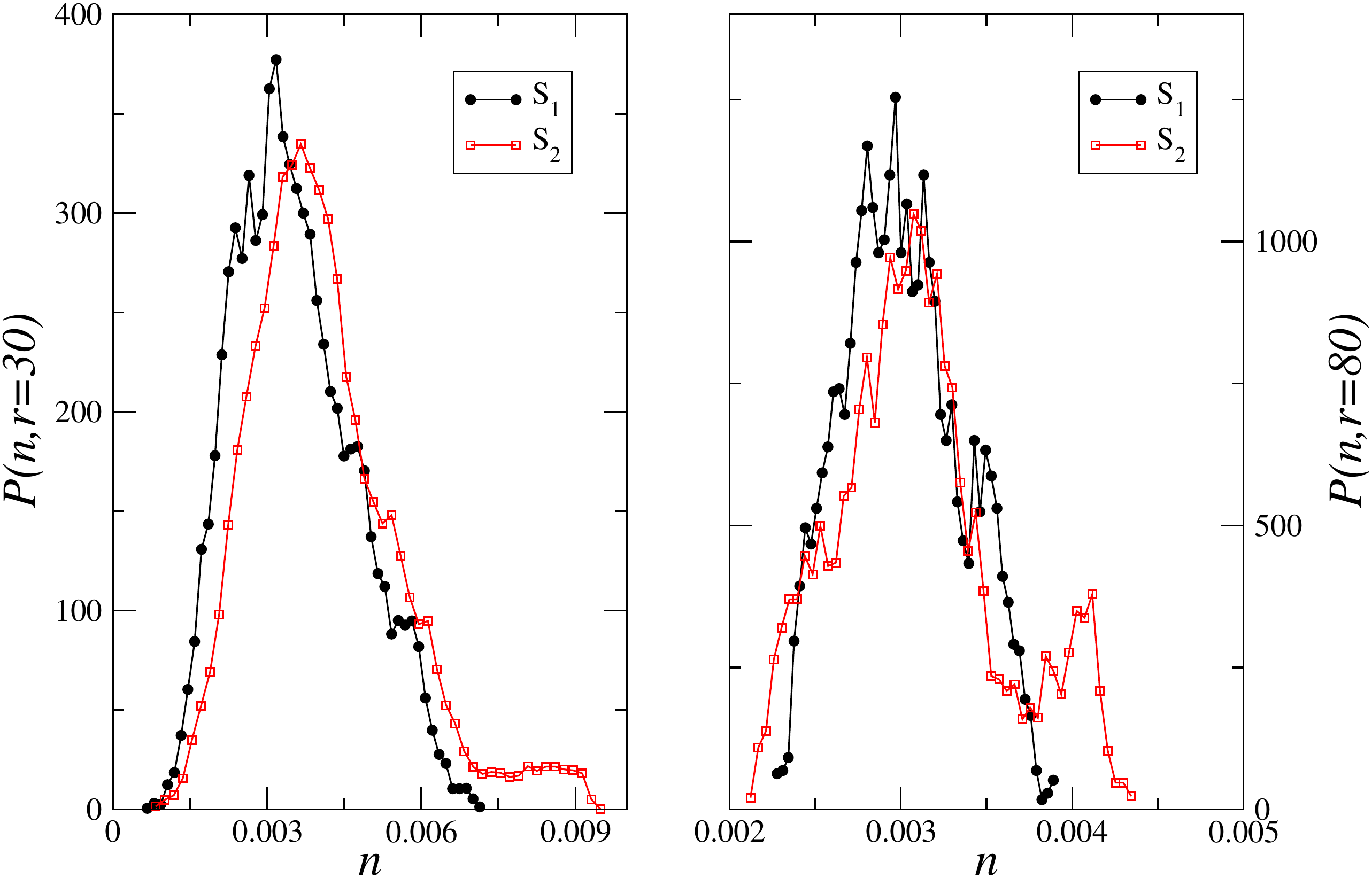}
\caption {PDFs of the conditional density in spheres of radius $r=30$
  Mpc/h (left) and $r=80$ Mpc/h (right), in two distinct regions: a
  nearby ($S_1$) and a faraway ($S_2$) one. Notice that the two PDFs
  statistically give the same signal.}
\label{fig:ss}
\end{figure}

Conditional fluctuations have been found to be not self-averaging in several SDSS-DR6 samples, {\it i.e.,} 
there were systematic differences between statistical properties measured in different parts of a given sample \cite{2df_epl,sdss_epl}. It was concluded that this
behavior is due to galaxy density fluctuations which are too large in
amplitude and too extended in space to be self-averaging inside the
considered volumes. The lack of self averaging prevents one to extract
a statistically meaningful information from whole sample average
quantities, as for example the conditional average density. 
We repeated the stability test of statistical quantities within the new SDSS-DR7
sample, since it almost doubled in size compared to SDSS-DR6. 
To this aim we cut the sample volume into two regions, a
nearby and a faraway one as in \cite{sdss_epl,sdss_aea}, and we
determine the PDFs $P(n(r))\equiv P(n;r)$ of the conditional density
separately in both regions, and at two different $r$ scales.  We
conclude from Fig.\ref{fig:ss} that the PDF is statistically stable
and does not show systematic dependence on system size, as opposed to
the case of the SDSS-DR6 data on scales $r>30$ Mpc/h
\cite{sdss_epl,sdss_aea}. Hence in this new sample, conditional
statistical quantities computed over the whole sample volume are
useful and meaningful indicators.

%%%%%%%%%%%%%%%%%%%%%
%%%%%%%%%%%%%%%%%%%%%

\subsection{Scaling at small scales} 

At small length scales ($r < 20$ Mpc/h) the exponent for the
conditional average density is close to minus one (see
Fig.\ref{fig:condden}).  This result is in agreement with ones
obtained by the same method in a number of different samples (see
\cite{sdss_epl,sdss_aea,2df_epl,2df_aea} and references therein).
This scaling can be interpreted as a signature of fractality of the
galaxy distribution in this range of scale. In addition, this implies
that the distribution is not uniform at these scales, and thus the
standard two-point correlation function is substantially biased.

%%%%%%
\subsection{Scaling at large scales}

\begin{figure}
\onefigure[scale=0.35]{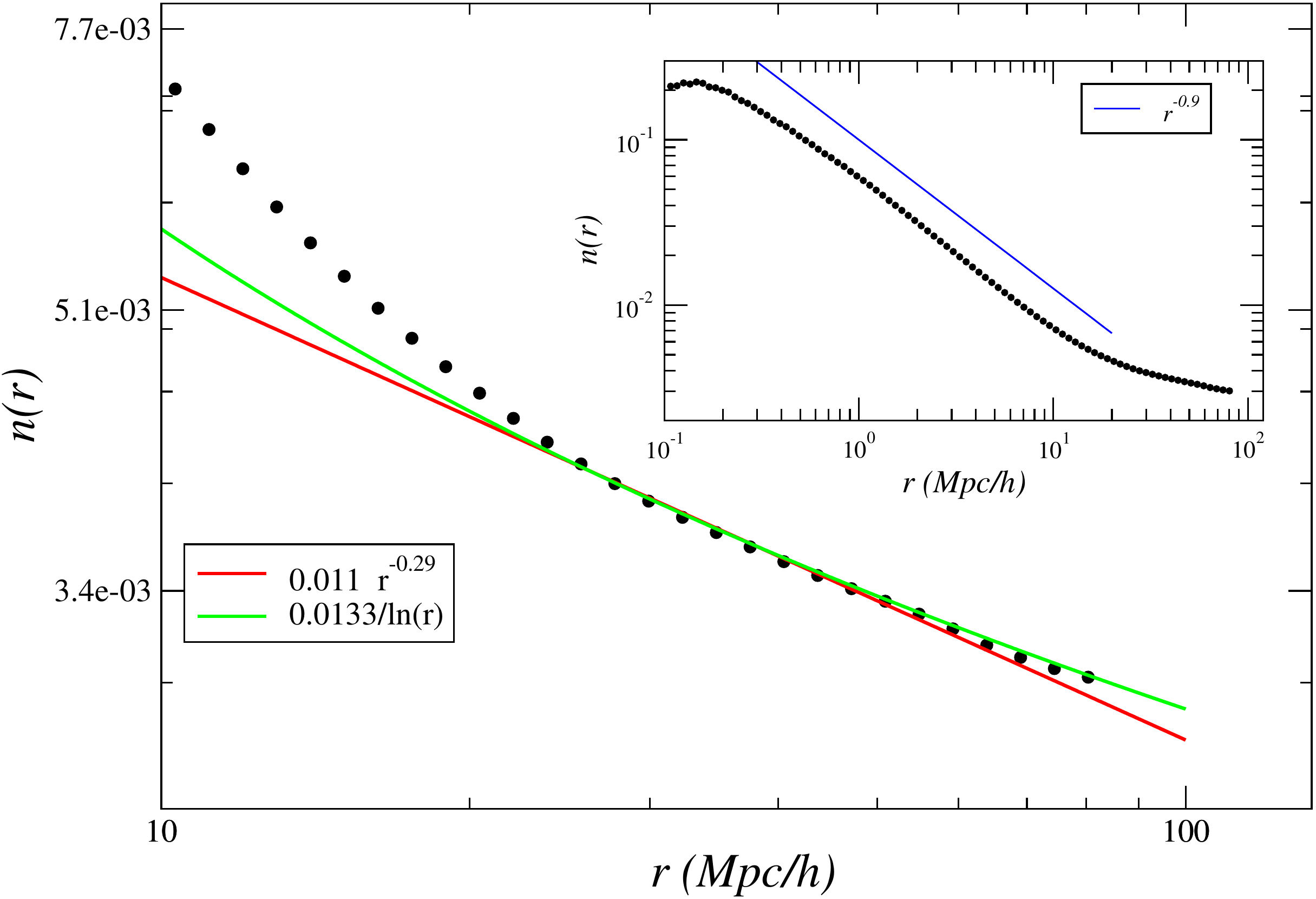}
\caption{Conditional average density $\overline n(r)$ of galaxies as a
  function of radius. In the inset panel the same is shown 
  in the full range of scales. Note the change of slope at $\approx
  20$ Mpc/h and also the lack of flattening up to $\approx
  80$ Mpc/h.  Our conjecture is that we
  have a logarithmic correction to the constant behavior, although we
  cannot exclude the possibility that it is power law with an exponent
  $\approx -0.3$.}
\label{fig:condden}
\end{figure}

We first computed the average conditional density
(Eq.\ref{eq:condden}) at large scales ($r > 10$ Mpc/h ). For a uniform
point distribution this quantity is constant, {\it i.e.,} independent
of the radius $r$ \cite{book}. Conversely, in our data we find a
pronounced $r$ dependence, as can be seen in
Fig.\ref{fig:condden}. Our best fit is
\begin{equation}
\label{densscale}
\overline{n(r)} \approx \frac{0.0133}{\log r} \;,
\end{equation}
that is the average density depends only weakly (logarithmically) on
$r$.  Alternatively, an almost indistinguishable power-law fit is
provided by
\begin{equation}
\label{densscale2}
\overline{n(r)} \approx 0.011 \times r^{-0.29} \;.
\end{equation}
We emphasize our preference for the logarithmic fit, where the only
fitting parameter is the amplitude. As can be seen in
Fig.\ref{fig:condden}, we find a change of slope in the conditional
average density in terms of the radius $r$ at about $\approx 20$
Mpc/h. At this point the decay of the density changes from an inverse
linear decay to a slow logarithmic one. Moreover, the density
$\overline{n(r)}$ does not saturate to a constant up to $\sim 80$
Mpc/h, {\it i.e.,} up to the largest scales probed in this sample.
Note that up to $r=80$ Mpc/h the number of points $M(r)$ is larger
than $10^4$, making this statistics very robust.

%%%%%%%%%%%%%%
%%%%%%%%%%%%%%

%%%%%%%%%%%%%%
\begin{figure}
\onefigure[scale=0.35]{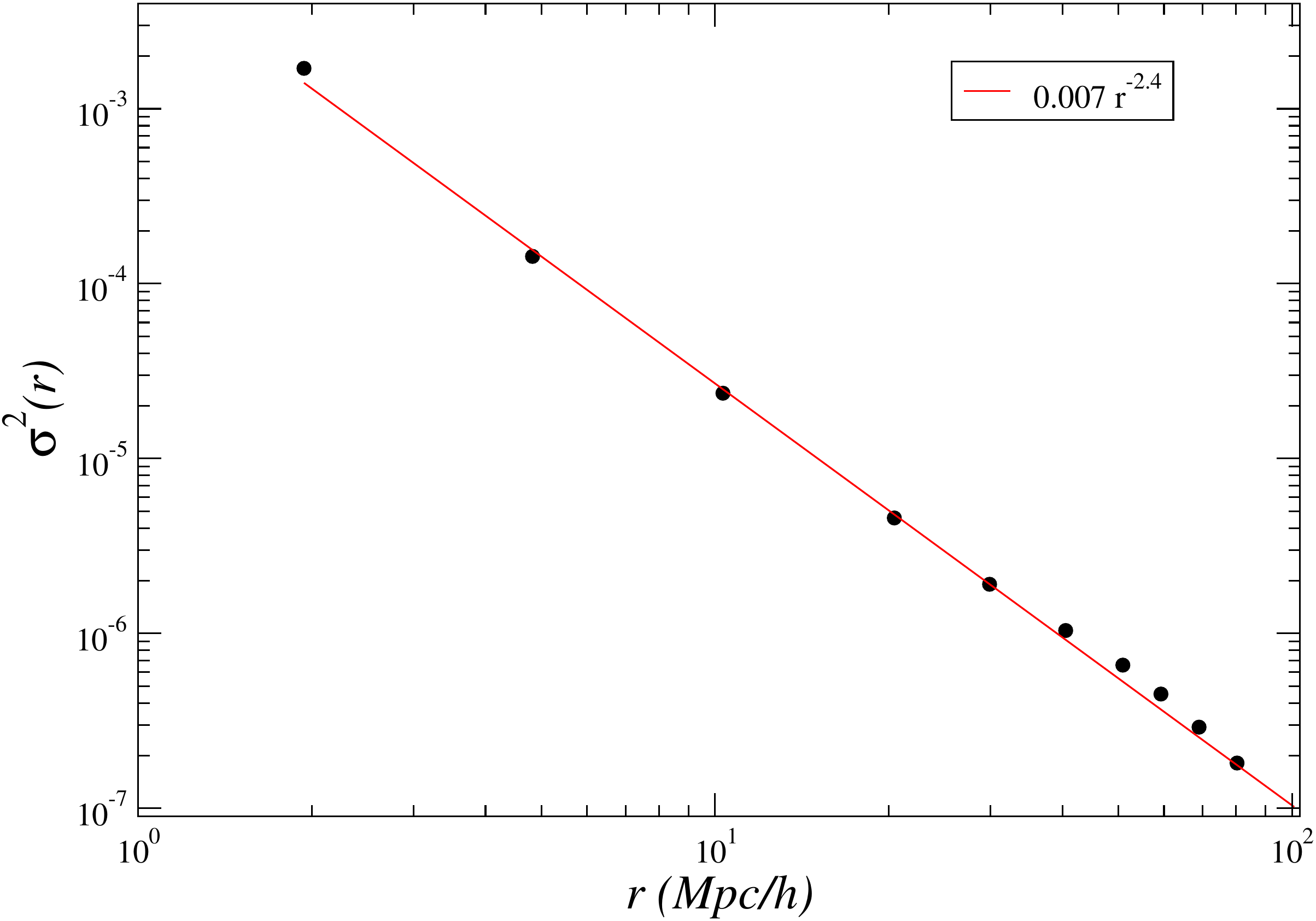}
\caption {Variance $\sigma^2$ of the conditional density $n_i(r)$ as a function of the radius.
Conversely, the corresponding variance of a Poisson point process would display a $1/r^3$ decay.}
\label{fig:variance}
\end{figure}
%%%%%%%%%%%%%%

This result is in agreement with a study of the SDSS-DR4 samples
\cite{dr4}, where, in the average conditional density, a similar
change of slope was observed at about the same scale $r \approx 20$
Mpc/h, together with quite large sample to sample fluctuations.
Indeed, some evidences were subsequently found to support that the
galaxy distribution is still characterized by rather large
fluctuations up to 100 Mpc/h, making it incompatible with uniformity
\cite{2df_epl,2df_aea,sdss_epl,sdss_aea,sdss_bao}.  Similarly, in the
Luminous Red Galaxy (LRG) sample of SDSS, Hogg et al. \cite{hogg} also
found a slope change in the average conditional density. On the other
hand, we do not observe a transition to uniformity at about 70 Mpc/h,
which they reported.  Note also that a study of the self-averaging
properties of fluctuations in the LRG sample is still lacking.

Compared to the average density, it is harder to find the correct fit
for the variance $\sigma^2(r)$ of the conditional density
(Eq.\ref{eq:condvar}).  Our best fit is (see Fig.\ref{fig:variance})
\begin{equation}
\label{denssigma2}
\sigma^2(r) \approx 0.007\times r^{-2.4} \;.
\end{equation}
Given the scaling behavior of the conditional density and variance, we
conclude that galaxy structures are characterized by non-trivial
correlations for scales up to $r \approx 80$ Mpc/h.

%%%%%%%%%%%%
\begin{figure}
\centering
\onefigure[scale=0.35]{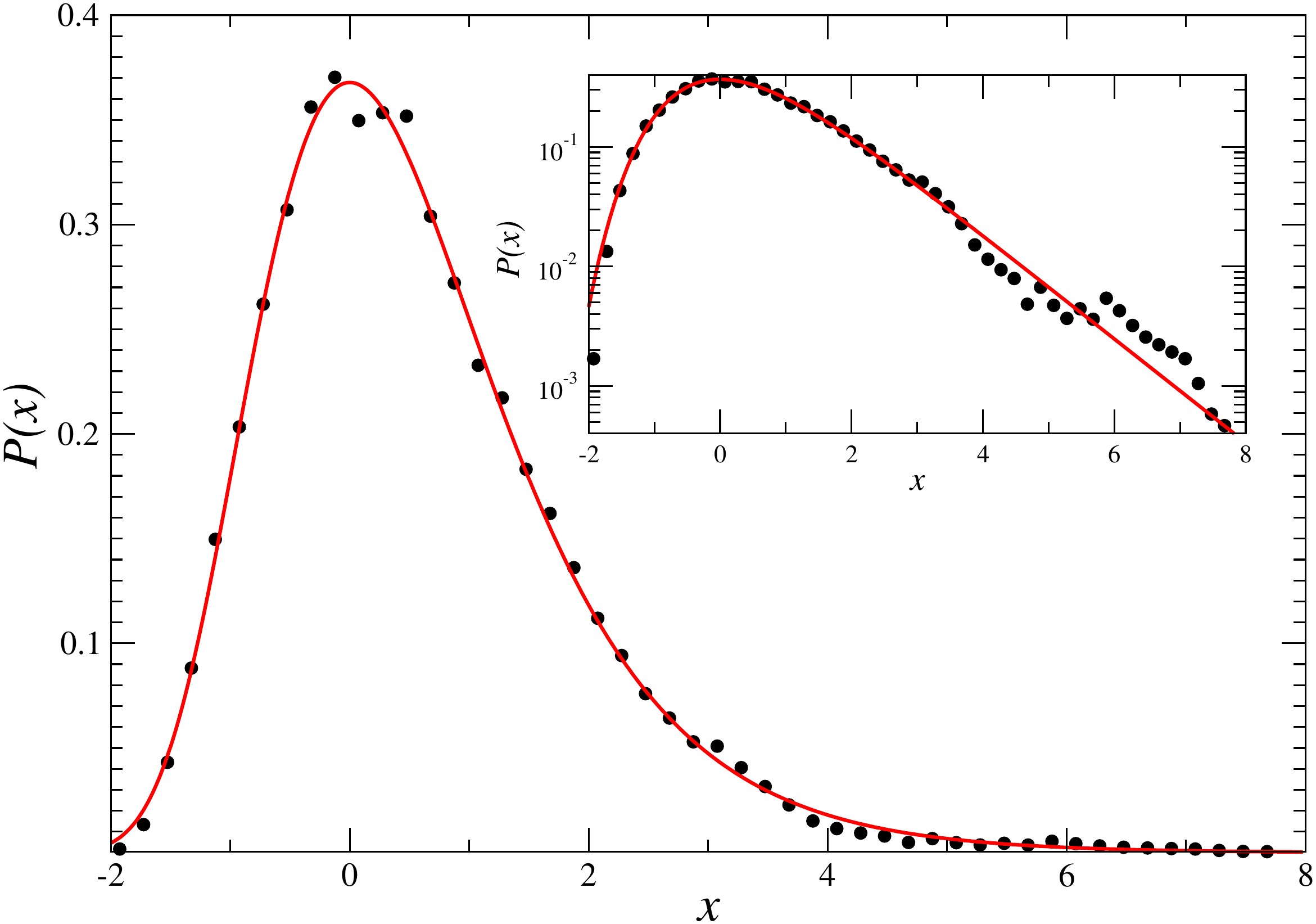}
\caption{One of the best fits is obtained for $r=20$. The data is
  rescaled by the fitted parameters $\alpha$ and $\beta$. The solid
  line corresponds to the parameter-less Gumbel distribution
  Eq.\ref{gumbelscaled}. The inset depicts the same on log-linear
  scale.}
\label{fig:best20}
\end{figure}
%%%%%%%%%%%%

\begin{figure}
\centering
\onefigure[scale=0.35]{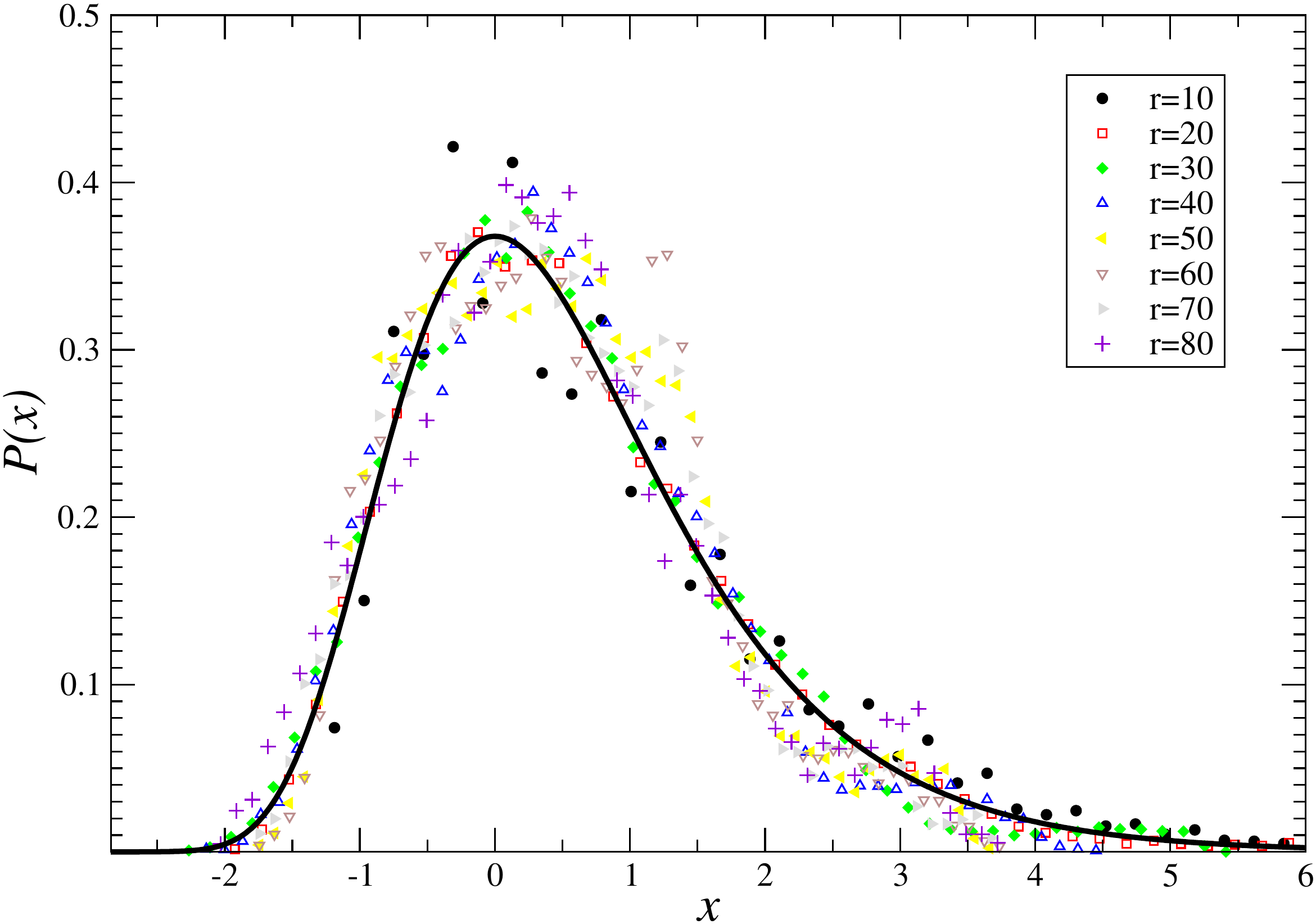}
\caption{Data curves of different $r$ scaled together by fitting
  parameters $\alpha$ and $\beta$ for each curves. The solid line is
  the parameter-free Gumbel distribution Eq.\ref{gumbelscaled}.}
\label{fig:scale}
\end{figure}

\begin{figure}
\centering
\onefigure[scale=0.35]{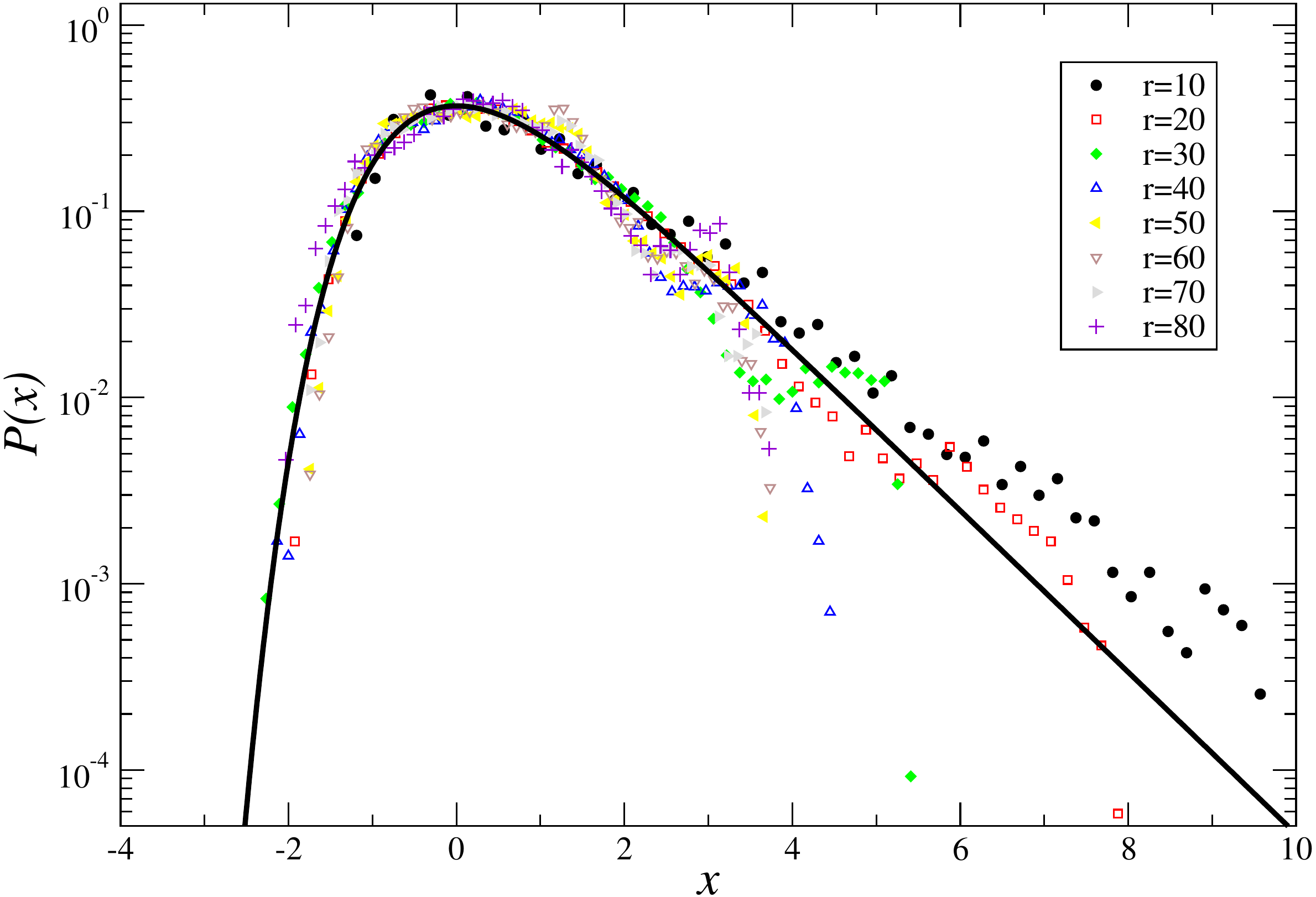}
\caption{The same as Fig.~\ref{fig:scale}, but on log-linear scale to
  emphasize the tails of the distribution.}
\label{fig:scalelog}
\end{figure}

To probe the whole distribution of the conditional density $n_i(r)$,
we fitted the Gumbel distribution (Eq.\ref{gumbel}) via its two
parameters $\alpha$ and $\beta$. One of our best fits is
  obtained for $r=20$ Mpc/h, see Fig.~\ref{fig:best20}. The data,
  moreover, convincingly collapses to the parameter-less Gumbel
  distribution (Eq.\ref{gumbelscaled}) for all values of $r$ for
  $10\le r\le 80$ Mpc/h, with the use of the scaling variable $x$ from
  Eq.\ref{scalevar} (see Figs.~\ref{fig:scale}-\ref{fig:scalelog}).
  Note that for a Poisson point process (uncorrelated random points)
  the number $N(r)$ (and consequently also the density) fluctuations
  are distributed exactly according to a Poisson distribution, which
  in turn converges to a Gaussian distribution for large average
  number of points $\overline{N(r)}$ per sphere. In our samples,
  $\overline{N(r)}$ is always larger than 20 galaxies, where the
  Poisson and the Gaussian PDFs differ less than the uncertainty in
  our data. Note also that due to the central limit theorem, all
  homogeneous point distributions (not only the Poisson process) lead
  to Gaussian fluctuations.  Hence the appearance of the Gumbel
  distribution is a clear sign of inhomogeneity and large scale
  structures in our samples.

The fitting parameters in Eq.\ref{gumbel} varied with the radius $r$
approximately as
\begin{equation}
\alpha \approx \frac{0.007}{r^{0.21}}, \quad \beta \approx
\frac{0.035}{r} \;.
\end{equation}
although a logarithmic fit $\alpha\approx 0.0115/\log r$ cannot be
excluded either.  With the fitted values of $\alpha$ and $\beta$ we
recover the (directly measured) average conditional density of
galaxies through Eq.~\ref{cumu}. On the other hand, we have a
discrepancy when comparing the directly measured $\sigma^2$ to that
obtained from the Gumbel fits through Eq.~\ref{cumu}. The reason for
this discrepancy is that the uncertainty in the tail of the PDF $P(n,r)$ is amplified when we directly calculate the second
moment.

%%%%%%%%%%%%
%%%%%%%%%%%%

%%%%%%%%%%%%
%%%%%%%%%%%%

%%%%%%%%%%%%%%%%%%%%%%%
\subsection{Data collapse without fitting}

\begin{figure}
\centering
\onefigure[scale=0.35]{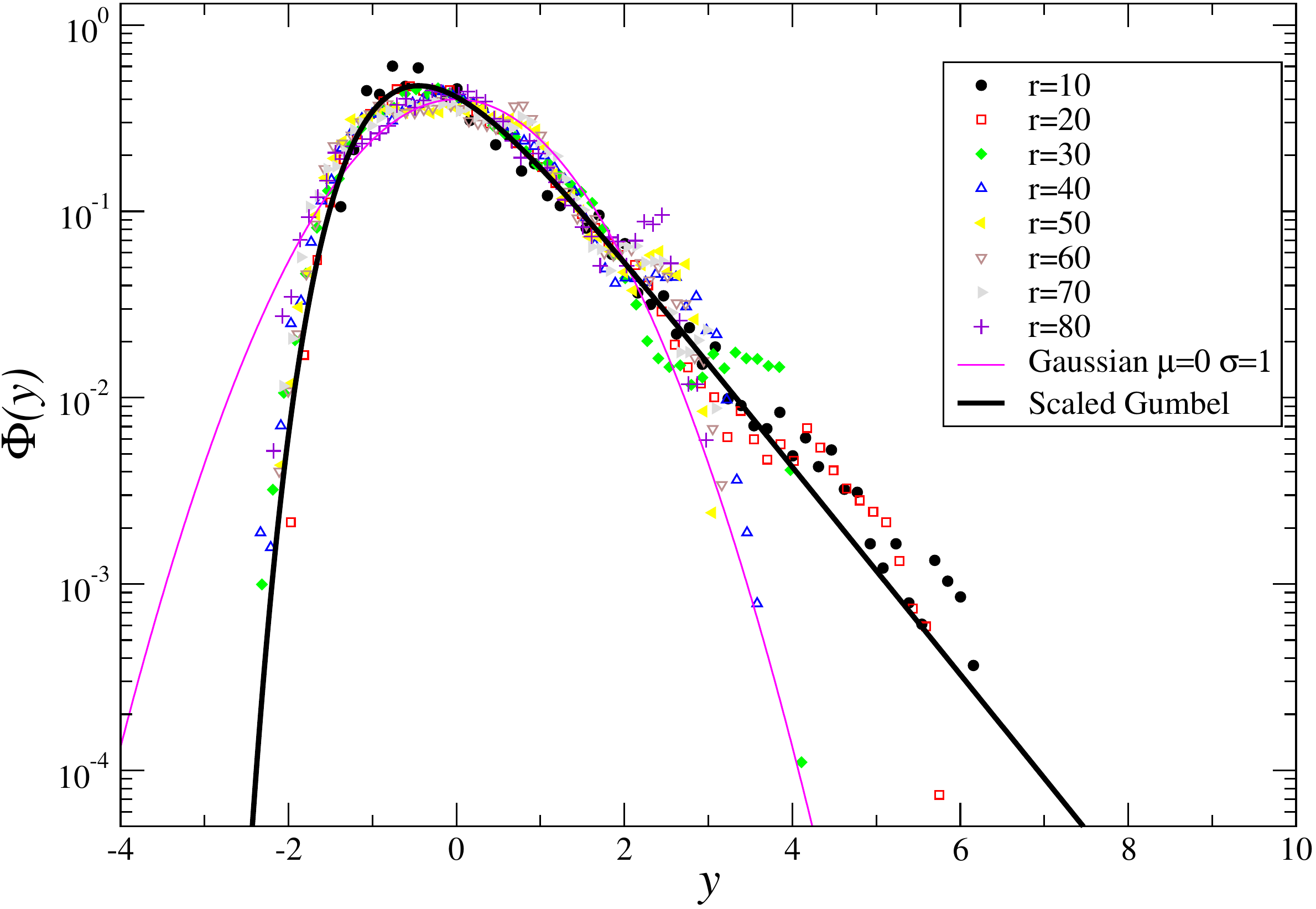} 
\caption{ The fitting-free data collapse
  (Eqs.~\ref{fittingfree1}-\ref{fittingfree2}) based on the first two
  moments of the distribution. Note again the satisfactory data
  collapse on all scales. The black line is the Gumbel distribution of
  Eq.(\ref{fittingfree2}), while the blue line is the corresponding
  normal Gauss distribution with zero mean and unit variance.}
\label{fig:scaled}
\end{figure}

It is possible to obtain a scaling of the data without any fitting
procedure. We can compute the average, $\mu$, and the standard
deviation, $\sigma^2$, of the data and use the scaled variable
\begin{equation}
\label{fittingfree1}
 y = \frac{N-\mu}{\sigma} \;.
\end{equation}
The density functions for different values of $r$ scale to the single curve
\begin{equation}
\label{fittingfree2}
 \Phi(y) = a e^{-(ay+\gamma)- e^{-(ax+\gamma)}}
\end{equation}
with $a=\pi/\sqrt 6$. (This function, of course, has mean zero, and
standard deviation one.)  This type of fitting-free data collapse has
been used extensively in statistical physics \cite{foltin94,antal01}.
As shown in Fig.\ref{fig:scaled} we find a satisfactory agreement with
Eq.\ref{fittingfree2}. Note also that Gaussian fluctuations can be
clearly excluded.  Compared to the fitting results of
Fig.~\ref{fig:scale}, the agreement in Fig.\ref{fig:scaled} is better
around the tails of the distribution, but it gets worse around the
maximum.  The reason for this latter mismatch is again due to the
uncertainty in the second moment.

%%%%%%%%%%%%%%%%%%%%%%%%%
\section{Discussion}

Given the observed scaling and data collapse in the spatial galaxy
data, is there any supporting evidence for the appearance of the
Gumbel distribution?  Due to the scaling and data collapse we argue
that the large scale galaxy distribution shows similarities with
critical systems.  Here the galaxy density around each galaxy is
analogous to a random variable describing a spatially averaged
quantity in a volume. The average conditional galaxy density depends
on the volume size ($\sim r^3$) only logarithmically
$\overline{n(r)}\sim 1/ \log r$ from Eq.\ref{densscale}. According to
the conjecture of Bramwell for critical systems \cite{bramwell09}, if
a spatially averaged quantity depends only weakly (say logarithmically) on
the system size, the distribution of this quantity follows the Gumbel
distribution. This is indeed what we see in the galaxy data. Hence our
two observations about the average density and the density
distribution are compatible with the behavior of critical systems in
statistical physics.

% We note that standard models of galaxy formation predict the matter
% density fields in the early universe to be weakly and long range
% correlated, {\it i.e.,} uniform. Gravitational clustering in an
% expanding universe, using such theoretical density field as initial
% conditions, only leads to small scale fluctuations in the matter
% distribution, and uniformity beyond $\approx 10$ Mpc/h
% \cite{cdm_theo,sdss_epl,sdss_aea}. Hence these theories cannot yet
% explain the non-Gaussian fluctuations we observed up to much larger
% scales (see \cite{cdm_theo,sdss_epl,sdss_aea,2df_epl,2df_aea} for
% more details).

We note that standard models of galaxy formation predict homogeneous
mass distribution beyond $\approx 10$ Mpc/h
\cite{cdm_theo,sdss_epl,sdss_aea}.  To explain our findings about
non-Gaussian fluctuations up to much larger scales presents a
challenge for future theoretical galaxy formation models (see
\cite{cdm_theo,sdss_epl,sdss_aea,2df_epl,2df_aea} for more details).

In summary, we have established scaling and data collapse over a wide
range of radius (volume) in galaxy data. Scaling in the data indicates
criticality. The average galaxy density depends only logarithmically
on the radius, which suggests a Gumbel scaling function
\cite{bramwell09}. The scaled data is indeed remarkably close to the
Gumbel distribution, which is one of the three extreme value
distributions. How this distribution arises through galaxy formation,
or what the extreme quantity is in the galaxy data, are challenging
questions needed to be addressed in the future.

\acknowledgments We are grateful to Andrea Gabrielli and Michael
Joyce, Luciano Pietronero, and Zolt\'an R\'acz for fruitful
discussions and valuable comments. TA acknowledges financial support
by the Templeton Foundation, the NSF/NIH Grant R01GM078986, the
Hungarian Academy of Sciences (OTKA No. K68109), and J. Epstein.  YVB
thanks for partial support from Russian Federation grants: Leading
Scientific School 1318.2008.2 and RFBR 09-02-00143.  We acknowledge
the use of the Sloan Digital Sky Survey data ({\tt
  http://www.sdss.org}) and of the NYU Value-Added Galaxy Catalog
({\tt http://ssds.physics.nyu.edu/}).

\end{document}